\documentclass[useAMS,usenatbib]{mn2e}
\usepackage{epsfig,color}
\def\reff@jnl#1{{\rm#1\/}}

\def\aj{\reff@jnl{AJ}}                  
\def\araa{\reff@jnl{ARA\&A}}            
\def\apj{\reff@jnl{ApJ}}                
\def\apjl{\reff@jnl{ApJ}}               
\def\apjs{\reff@jnl{ApJS}}              
\def\apss{\reff@jnl{Ap\&SS}}            
\def\aap{\reff@jnl{A\&A}}               
\def\aapr{\reff@jnl{A\&A~Rev.}}         
\def\aaps{\reff@jnl{A\&AS}}             
\def\baas{\reff@jnl{BAAS}}              
\def\jrasc{\reff@jnl{JRASC}}            
\def\memras{\reff@jnl{MmRAS}}           
\def\mnras{\reff@jnl{MNRAS}}            
\def\physrep{\reff@jnl{Phys.Rep.}}
\def\pra{\reff@jnl{Phys.Rev.A}}         
\def\prb{\reff@jnl{Phys.Rev.B}}         
\def\prc{\reff@jnl{Phys.Rev.C}}         
\def\prd{\reff@jnl{Phys.Rev.D}}         
\def\prl{\reff@jnl{Phys.Rev.Lett}}      
\def\pasp{\reff@jnl{PASP}}              
\def\pasj{\reff@jnl{PASJ}}              
\def\skytel{\reff@jnl{S\&T}}            
\def\solphys{\reff@jnl{Solar~Phys.}}    
\def\sovast{\reff@jnl{Soviet~Ast.}}     
\def\ssr{\reff@jnl{Space~Sci.Rev.}}     
\def\nat{\reff@jnl{Nature}}             

\begin{document}

\title[Autocorrelations of  stellar light and  mass]
{Autocorrelations of stellar light and  mass at
$z\sim0$ and  $\sim1$: From  SDSS to  DEEP2}

\author[C. Li et al.]
{Cheng Li$^{1,2}$\thanks{E-mail: leech@shao.ac.cn},Simon D.~M. White$^{2}$,
Yanmei Chen$^{3}$,Alison L. Coil$^{4}$\thanks{Alfred P. Sloan Foundation Fellow},
Marc Davis$^{5}$,\newauthor
Gabriella De Lucia$^{6}$,Qi Guo$^{7}$,Y. P. Jing$^{1}$,Guinevere Kauffmann$^{2}$,
Christopher N. \newauthor
A. Willmer$^{8}$ and Wei Zhang$^{9}$ \\
$^{1}$Partner Group of the MPI f\"{u}r Astrophysik
at Shanghai Astronomical Observatory,
Key Laboratory for Research in Galaxies \\ 
~and Cosmology of Chinese Academy of Sciences,
Nandan Road 80, Shanghai 200030, China\\
$^{2}$Max-Planck-Institute f\"{u}r Astrophysik,
Karl-Schwarzschild-Str. 1, D-85741 Garching, Germany\\
$^{3}$Department of Astronomy, Nanjing University, Nanjing 210093, China \\
~Key Laboratory of Modern Astronomy and  Astrophysics (Nanjing University), 
Ministry of Education, Nanjing 210093, China \\
$^{4}$Department of Physics, Center for Astrophysics and Space Sciences, 
University of California, \\
~9500 Gilman Dr., La Jolla, San Diego, CA 92093, USA\\
$^{5}$Department of Astronomy, University of California, Berkeley, CA 94720, USA\\
$^{6}$INAF - Astronomical Observatory of Trieste, via G.B. Tiepolo 11,
I-34143 Trieste, Italy\\
$^{7}$Institute for Computational Cosmology, Department of Physics, 
University of Durham, South Road, Durham, DH1 3LE, UK\\
$^{8}$Steward Observatory, University of Arizona, Tucson, AZ 85721, USA\\
$^{9}$National Astronomical Observatories, Chinese Academy of Sciences, 
Beijing 100012, China
}

\date{Accepted ........ Received ........; in original form ........}

\pagerange{\pageref{firstpage}--\pageref{lastpage}} \pubyear{2010}

\maketitle

\label{firstpage}

\begin{abstract}
We  present   measurements  of  projected   autocorrelation  functions
$w_p(r_p)$ for the stellar mass of galaxies and for their light in the
$U$, $B$ and $V$ bands, using  data from the third data release of the
DEEP2 Galaxy Redshift  Survey and the final data  release of the Sloan
Digital  Sky Survey  (SDSS).  We investigate  the  clustering bias  of
stellar   mass   and   light   by   comparing   these   to   projected
autocorrelations  of   dark  matter  estimated   from  the  Millennium
Simulations  (MS) at  $z=1$ and  $0.07$, the  median redshifts  of our
galaxy samples.   All of the  autocorrelation and bias  functions show
systematic  trends   with  spatial   scale  and  waveband   which  are
impressively  similar  at  the  two  redshifts. This  shows  that  the
well-established  environmental dependence  of stellar  populations in
the local Universe  is already in place at  $z=1$. The recent MS-based
galaxy   formation  simulation   of   \citet{Guo-11}  reproduces   the
scale-dependent clustering  of luminosity  to an accuracy  better than
30\%  in   all  bands  and   at  both  redshifts,   but  substantially
overpredicts mass  autocorrelations at separations below  about 2 Mpc.
Further  comparison of  the  {\it  shapes} of  our  stellar mass  bias
functions with  those predicted  by the model  suggests that  both the
SDSS and  DEEP2 data prefer  a fluctuation amplitude  of $\sigma_8\sim
0.8$ rather than the $\sigma_8 = 0.9$ assumed by the MS.
\end{abstract}

\begin{keywords}
galaxies: clusters:  general --  galaxies: distances and  redshifts --
cosmology: theory -- dark matter -- large-scale structure of Universe.
\end{keywords}

\section{Introduction}\label{sec:introduction}

The  two-point correlation  function  (2PCF) has  long  served as  the
primary way of quantifying the spatial distribution of galaxies in our
Universe.  Measurements  of 2PCF for different classes  of galaxies in
the local Universe have been  carried out with high accuracy thanks to
the large  redshift surveys assembled  in recent years,  in particular
the       two-degree      field      galaxy       redshift      survey
\citep[2dFGRS;][]{Colless-01}  and   the  Sloan  Digital   Sky  Survey
\citep[SDSS;][]{York-00}.  These  studies have established  that 2PCFs
depend on  a variety of  properties such as luminosity,  stellar mass,
colour,  spectral type  and  morphology \citep{Davis-88,  Hamilton-88,
  White-Tully-Davis-88,   Boerner-Mo-Zhou-89,   Einasto-91,   Park-94,
  Loveday-95,   Benoist-96,  Guzzo-97,  Willmer-daCosta-Pellegrini-98,
  Loveday-Tresse-Maddox-99,                       Beisbart-Kerscher-00,
  Brown-Webster-Boyle-00, Guzzo-00, Norberg-01, Norberg-02, Zehavi-02,
  Zehavi-05,   Li-06b,  Wang-07b,   Swanson-08,    Wang-10,
  Ross-Tojeiro-Percival-11, Zehavi-11}.  Recently,  there have also  
been studies
of galaxy clustering  at higher redshifts, which are  usually based on
deep  surveys  (up to  $z\sim1-1.5$)  covering  small areas  ($\la$1.5
deg$^{2}$),  e.g.  the DEEP2  Galaxy Redshift  Survey \citep{Davis-03,
  Coil-04a,   Coil-06a,   Coil-08a},   the   VIMOS-VLT   Deep   Survey
\citep{LeFevre-05, Pollo-05, Pollo-06,  Meneux-06, Meneux-08}, and the
zCOSMOS  Survey \citep{Lilly-07, Lilly-09,  Meneux-09, delaTorre-11a}.
Measurements  of  2PCFs  for  different  classes of  galaxies  and  at
different redshifts have provided powerful quantitative constraints on
models   of   galaxy    formation   and   evolution   \citep[see   for
  example][]{DeLucia-Blaizot-07, Li-07a, Guo-11, delaTorre-11b}.

A number of recent studies have further investigated the dependence of
galaxy clustering  on physical properties by  measuring {\em weighted}
or    {\em    marked}    2PCFs    \citep[e.g.][]{Beisbart-Kerscher-00,
  Faltenbacher-02,  Sheth-05,  Skibba-06, Skibba-Sheth-09,  Skibba-09,
  Li-White-09,  Li-White-10}.   Simply  speaking,  this  statistic  is
estimated using exactly  the same methodology as the  one used for the
traditional 2PCF, except that each galaxy in the real sample and/or in
the random sample  is weighted by one of  its physical properties such
as stellar  mass or luminosity at  a given band. When  compared to the
traditional  2PCF,  this   alternative  two-point  statistic  has  the
advantage  that it  makes  use of  the  whole galaxy  sample and  thus
minimizes the sampling and  large-scale structure noises.  As found in
\citet{Li-White-09}, a  particular virtue of  the weighted correlation
functions   is  that   the  correlation   signals  are   dominated  by
contributions from galaxies in  a relatively narrow mass range, around
the  characteristic mass  of  the stellar  mass  function.  Thus,  the
statistic is  robust to incompleteness  of the observed sample  at the
two  mass  extremes.   In  addition,  the weighting  scheme  used  for
estimating the statistic  makes it a direct measure  of the clustering
of {\em  stars} on  scales larger than  those of  individual galaxies,
rather  than the  clustering of  the host  galaxies as  probed  by the
traditional 2PCF.  This statistic thus provides a compact and accurate
way to characterize the distribution of stellar populations over large
ranges in spatial scale.

Using  a sample  of  almost half  a  million galaxies  from the  Sloan
Digital  Sky Survey  (SDSS), \citet[][hereafter  Paper I]{Li-White-09}
and  \citet[][hereafter  Paper  II]{Li-White-10}  estimated  projected
autocorrelation  functions $w_p(r_p)$  both  for the  stellar mass  of
galaxies and for  their light in the five  SDSS photometric bands. All
of the autocorrelation estimates are extremely well described by power
laws        over        the        full        non-linear        range
$10h^{-1}$kpc$<r_p<10h^{-1}$Mpc. Luminosity  is found to  cluster less
strongly  than stellar  mass  in all  bands  and on  all scales.   The
autocorrelation  function  of  luminosity varies  systematically  with
wavelength in both amplitude and slope, with the reddest band (the $z$
band) showing the highest amplitude and the steepest slope, indicating
that the $z$-band light is the closest proxy for stellar mass in terms
of   clustering   properties.   These   trends   provide   a   precise
characterization of  the well-known dependence  of stellar populations
on  environment.   In combination  with  autocorrelation functions  of
luminosity-  and  stellar mass-selected  galaxy  samples,  as well  as
accurate luminosity and stellar  mass functions, these results provide
tight constraints on galaxy formation, which are a major challenge for
current  models (see,  for example,  the significant  discrepancies in
recent work by \citealt{Guo-11} based on the Millennium Simulations).

In  this paper,  we  extend the  work of  Papers  I and  II to  higher
redshifts ($z\sim1$) using data  from the DEEP2 Galaxy Redshift Survey
\citep{Davis-03}.   We use  the  methodology  of Papers  I  and II  to
compute  projected stellar  mass autocorrelation  functions  for DEEP2
galaxies, as  well as projected autocorrelations  for their luminosity
in  the rest-frame  $U$ and  $B$ bands.  In order  to compare  the two
surveys  directly,  we  re-analysize  the SDSS,  computing  luminosity
autocorrelations in the $U$, $B$  and $V$ bands, where luminosities in
these bands are estimated from  SDSS data through a Bayesian technique
based on  a spectral energy distribution library  constructed from the
\citet[][hereafter   BC03]{Bruzual-Charlot-03}   population  synthesis
code.   We compare  these results  with the  autocorrelations  of dark
matter at  $z=0.07$ and  $z=1$, the median  redshifts of the  SDSS and
DEEP2 galaxy samples used here,  in order to understand the {\em bias}
of  stellar  mass   and  light.   By  comparing  with   the  model  of
\citet{Guo-11},  we  investigate   quantitatively  how  well  current
treatments of  galaxy formation reproduce the  clustering evolution of
stellar mass and  light. Finally, we discuss the  possibility that the
measured shape  of the mass  autocorrelation functions can be  used to
estimate  the  value  of  the  mass  fluctuation  amplitude  parameter
$\sigma_8$.

\section{Data}

\subsection{SDSS galaxy sample}

The   low-redshift   galaxy  sample   used   in   this   study  is   a
magnitude-limited  sample  constructed  from  the final  data  release
\citep[DR7;][]{Abazajian-09}  of  the   SDSS  \citep{York-00}  and  is
exactly the same as that used in Papers I and II. This sample contains
482,755 galaxies located in the  main contiguous area of the survey in
the  northern Galactic cap,  with $r<17.6$,  $-24<M_{^{0.1}r}<-16$ and
spectroscopically measured redshifts in the range $0.001<z<0.5$.  Here
$r$  is  the  $r$-band  Petrosian apparent  magnitude,  corrected  for
Galactic  extinction,  and  $M_{^{0.1}r}$  is the  $r$-band  Petrosian
absolute magnitude,  corrected for evolution and  $K$-corrected to its
value at $z=0.1$.  The apparent  magnitude limit is chosen in order to
select a sample  that is uniform and complete over  the entire area of
the  survey \citep[see][]{Tegmark-04}.   The median  redshift  of this
sample is $z=0.088$, with 10\%  of the galaxies below z=0.033 and 10\%
above  z=0.16.    As  shown  in   Paper  I  (see  their   fig.~4)  the
autocorrelation of  stellar mass is dominated by  contributions from a
narrower and slightly lower redshift  range (with 10\% of the galaxies
below $z=0.025$  and 10\% above  $z=0.12$), with a median  redshift of
$z=0.067$.   We thus  take  $z=0.067$ as  the  {\em effective}  median
redshift   of   this   sample   when  comparing   with   dark   matter
autocorrelations and semi-analytical model predictions.

We  use a  Bayesian  technique  to derive  estimates  of the  absolute
magnitudes in  $U$, $B$ and $V$  bands for each galaxy  in our sample,
following  \citet{Kauffmann-03}  and  \citet{Salim-05}.  Libraries  of
Monte  Carlo  realizations  of  model  star  formation  histories  are
generated between $0<z<0.5$ in regular bins of $\Delta z=0.001$, using
the BC03 population synthesis code. Each library contains 25000 models
with  each  star  formation   history  being  characterized  with  two
components:  an  underlying  continuous  model with  an  exponentially
declining star  formation law and  random bursts superimposed  on this
continuous  model.   The  models  also  have  metallicities  and  dust
attenuation  uniformly distributed  over wide  ranges.   The universal
initial mass function (IMF)  of \citet{Kroupa-01} is adopted. For each
galaxy in  our sample, we  derive the $U$,  $B$ and $V$  magnitudes by
comparing  the observed  $ugriz$  SED to  all  the model  SEDs in  the
closest redshift model  library. The $\chi^2$ goodness of  fit of each
model  determines  the  weight,  $\propto \exp(-\chi^2/2)$,  which  is
assigned  to that  model  when building  the probability  distribution
functions (PDFs) of  the restframe $U$, $B$ and  $V$ magnitudes of the
galaxy. We adopt  the PDF-weighted {\em mean} values  as our estimates
of these quantities. Adopting the {\em median} of the PDF gives almost
identical results for the autocorrelation function analysis.

\subsection{DEEP2 galaxy sample}

The  DEEP2 Galaxy  Redshift Survey  \citep[][]{Davis-03}  utilizes the
DEIMOS spectrograph \citep{Faber-03} on the KECK II telescope. Targets
for  the  spectroscopic sample  were  selected  from $BRI$  photometry
\citep{Coil-04a}  taken  with  the  12k  x 8k  mosaic  camera  on  the
Canada-France-Hawaii  Telescope  (CFHT). The  images  have a  limiting
magnitude of $R_{\rm  AB} \sim 25.5$. Since the  $R$-band provides the
highest  signal-to-noise ratio  (S/N) among  all the  CFHT  bands, the
photometry in this  band was used to select  targets for spectroscopic
observation   in   the   DEEP2.   The   CFHT   imaging   covers   four
widely-separated regions,  with a  total area of  3.5 $\rm  deg^2$. In
fields 2 to  4, the spectroscopic sample is  preselected using ($B-R$)
and    ($R-I$)   colors    to   eliminate    objects    with   $z<0.7$
\citep{Davis-03}. Color and apparent  magnitude cuts were also applied
to objects  in the  first field, the  Extended Groth Strip  (EGS), but
these were  designed to downweight  low redshift galaxies  rather than
eliminate them entirely \citep{Willmer-06}.  

We   use   data  from   the   third   data   release  of   the   DEEP2
survey\footnote{http://deep.berkeley.edu/DR3/}  which contains spectra
of about  50000 galaxies in the  magnitude range $18.5  \le R_{\rm AB}
\le 24.1$. The  spectra have a resolution of $R  \sim 5000$.  For this
study we have selected a sample  of 30546 galaxies from the DEEP2 DR3,
with redshifts  of quality 3 or  4 and in the  range $0.8<z<1.35$. The
median redshift of this sample  is $z=1$, which we use for comparisons
with dark matter and model galaxy autocorrelations.

The derived  galaxy parameters required  in this work  include stellar
mass  ($M_*$) and  restframe magnitudes  in  $U$ and  $B$ bands.   The
procedure for estimating  these parameters is exactly the  same as the
one above  for the  SDSS.  In  brief, the observed  $BRI$ SED  of each
galaxy in  the DEEP2 is compared to  a large grid of  BC03 model SEDs,
providing a maximum likelihood  estimate of the $I$-band mass-to-light
ratio of the galaxy, as well  as its restframe $U$ and $B$ magnitudes.
A \citet{Kroupa-01} initial mass function  (IMF) is adopted, as in the
SDSS analysis. Our estimates  of stellar mass and restframe magnitudes
are statistically well consistent with those from \citet{Bundy-06} and
\citet{Willmer-06}.  Indeed  we have repeated  our clustering analysis
using  stellar  mass  and  restframe magnitude  estimates  from  these
previous papers, obtaining almost the same results.

We don't consider the $V$ band for DEEP2 galaxies, as at $z\sim1$ this
band is  shifted well beyond the  reddest band (the $I$  band) that is
observed.

\subsection{Semi-analytic model galaxy catalogues}

In this paper we compare our observational results to predictions from
the galaxy  formation model of  \citet[][hereafter G11]{Guo-11}. This
model was  created by  implementing semi-analytic models  for baryonic
astrophysics  on  merger  trees  encapsulating the  evolution  of  the
halo/subhalo  population in  the  Millennium \citep{Springel-05b}  and
Millennium-II   \citep{Boylan-Kolchin-09}  Simulations\footnote{Galaxy
  catalogues  for  this  model  and  halo catalogues  for  the  parent
  simulations         are         publicly        available         at
  http://www.mpa-garching.mpg.de/millennium}.    The   Millennium  was
carried out  in a  cubic region  500 $h^{-1}$Mpc on  a side  with mass
resolution  $\sim 10^9$  M$_\odot$, while  the  Millennium-II followed
evolution in a region with 125  times smaller volume, but at 125 times
better mass resolution. The  combination of the two simulations allows
galaxy  formation  to be  studied  over  the  full range  of  observed
populations, from dwarf spheroidals  to cD galaxies.  In comparison to
earlier semi-analytic models from  the Munich group, the treatments of
supernova  feedback,  galaxy  size,  photoionisation  suppression  and
environmental  effects on satellite  galaxies have  been significantly
updated, resulting in  excellent fits not only to  recent SDSS data on
the luminosity and stellar mass functions of galaxies, but also to the
recent  determinations of  the abundance  of faint  satellite galaxies
around the  Milky-Way. Most of the population  properties for galaxies
in the  local Universe  are reasonably well  reproduced by  the model,
which also makes precise predictions for nonlinear clustering over the
full observed range of  scales, 10~kpc to 10~Mpc.  Comparisons between
the clustering  properties predicted by the model  and measurements of
SDSS galaxy  clustering show  that agreement with  SDSS data  is quite
good for masses above $6\times10^{10}M_\odot$ and at separations above
2 Mpc,  although the predicted  clustering is up  to 20\% too  high in
some mass ranges.  On smaller scales lower mass galaxies are predicted
to be substantially more clustered than is observed.  G11 suggest, but
do  not prove,  that this  may  be a  consequence of  an overly  large
present-day fluctuation amplitude ($\sigma_8=0.9$) in the simulations.

\section{Clustering measures}

Following Papers  I and II,  we weight each  galaxy in our  samples to
correct  for  incompleteness  when  computing  our  stellar  mass  (or
luminosity) autocorrelations. The weights  used for SDSS galaxies take
into account three factors. The first is $1/f_{sp}$, where $f_{sp}$ is
a  spectroscopic   completeness,  defined  as  the   fraction  of  the
photometrically defined target galaxies  for which usable spectra were
obtained. The  second weight, $1/V_{ij}$,  is applied to each  pair of
galaxies, where  $V_{ij}=\min(V_{max,i},V_{max,j})$ and $V_{max,i}$ is
the maximum  volume over which the  $i$th galaxy would  be included in
the sample. This weight accounts  for the fact that faint galaxies are
not  detected throughout the  entire survey  volume in  a flux-limited
survey. The final weight is  the factor $1/f_{coll,ij}$, which is also
applied to  galaxy pairs  and appears in  data-data counts  only. This
factor is  a function of  angular separation $\theta_{ij}$ of  the two
galaxies  and corrects  for the  effect of  fibre collisions  on small
scales.

Similarly, for each  galaxy pair in the DEEP2 sample  we also assign a
weight following \citet{Willmer-06}:
\begin{equation}
    W_{ij} = \frac{\kappa_i\kappa_j}{V_{ij}},
\end{equation}
where $\kappa_i$ accounts for  incompleteness resulting from the DEEP2
colour selection and redshift success rate. The second factor $V_{ij}$
is  defined in  the same  way as  above. Detailed  description  on the
calculation of the weights can be found in \citet{Willmer-06}.

We  have generated  a random  sample which  has the  same  overall sky
coverage and  redshift distribution as  the DEEP2 sample.  The spatial
window function of  the DEEP2 survey is applied  to the random sample,
which  includes  masking areas  around  bright  stars  and takes  into
account  the  varying  redshift  completeness  of  the  observed  slit
masks.   The    reader   is    referred   to   previous    papers   by
\citet{Coil-04b,Coil-06a,Coil-08a}  for  detailed  description of  the
window function. For each real galaxy we generate 100 sky positions at
random within the entire survey  region of the DEEP2 including all the
four separate fields, and we assign  to each of them the properties of
the  real   galaxy,  in  particular,   its  values  of   $\kappa$  and
$V_{max}$. Since the  four fields of the survey  are widely separated,
correlations in these  properties in the real sample  are wiped out by
randomizing in angle. We  follow \citet{Coil-06a} to assign a redshift
to each  random point according to the  redshift distribution averaged
over all the fields in the data.

\begin{figure*}
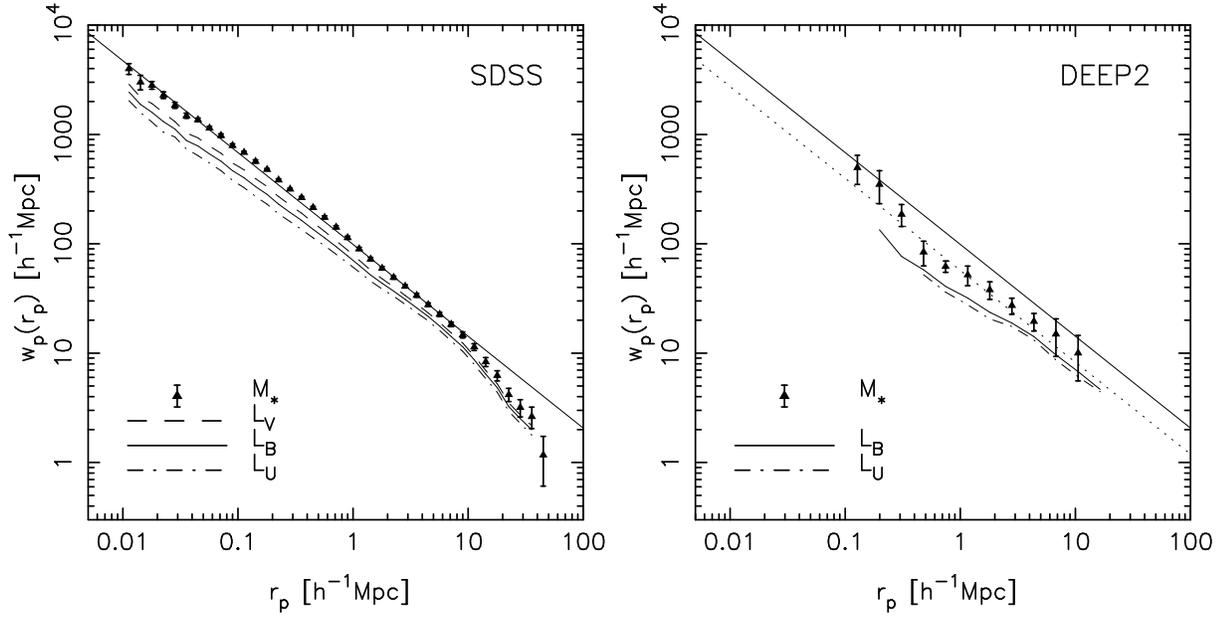

  \centerline{
    \epsfig{figure=f1a.ps,width=0.45\textwidth}
    \epsfig{figure=f1b.ps,width=0.45\textwidth} }
  \caption{Projected  stellar mass  autocorrelation  functions in  the
    SDSS/DR7  (left panel)  and  in the  DEEP2/DR3  (right panel)  are
    plotted as triangles with error bars and are compared to projected
    luminosity  autocorrelation functions  measured in  the rest-frame
    $U$,  $B$ and  $V$  bands (  lines).  Errors on  the stellar  mass
    autocorrelation function are estimated  in the left panel from the
    scatter among measurements on  20 mock SDSS catalogues constructed
    from the  Millennium Simulation, and  in the right panel  from the
    scatter between measurements on the four independent fields of the
    DEEP2 survey.  The solid line in the left panel is a power-law fit
    to the SDSS data  over the range $10h^{-1}kpc<r_p<10h^{-1}Mpc$ and
    is   repeated  in   the   right  panel.   It   corresponds  to   a
    three-dimensional             autocorrelation             function
    $\xi^\ast(r)=(r/r_0)^{-1.84}$ with $r_0=6.1h^{-1}Mpc$.  The dotted
    line in the right panel is  a power-law fit to the DEEP2 data over
    the full range probed, with index fixed to be -1.84, i.e. the same
    as  what  is determined  for  the  SDSS  data.  The  corresponding
    correlation length is $r_0=4.3h^{-1}Mpc$.}
  \label{fig:wrp}
\end{figure*}

The projected  autocorrelation function of stellar  mass or luminosity
in  a given band  is then  computed using  the estimator  described in
Paper I  (see their eqn. 2),  in which the weights  obtained above are
applied. We  estimate the  autocorrelation functions using  the entire
real (or random) sample, for both  SDSS and DEEP2.  Errors on the mass
autocorrelation in the  SDSS are estimated from the  scatter among the
measurements  from  20 mock  galaxy  catalogues  constructed from  the
Millennium Simulation  using the same  selection criteria as  the real
sample  (see Paper  I  for details).   The  errors on  the DEEP2  mass
autocorrelation function come from  the scatter among the measurements
for  the four  separate fields  of  the survey.   These errors  should
include  both the effect  of counting  noise and  that of  {\em cosmic
  variance}, which are impressively small in the SDSS due to the large
size and volume  of the sample.  Following Paper I,  we do not attempt
to   put  independent   error   bars  on   the  projected   luminosity
autocorrelations because,  for a given survey, with  our technique the
set of galaxy  pairs used to estimate each of  these functions is {\em
  exactly}  the same and  as a  result the  noise fluctuations  due to
sampling   and  to   large-scale  structure   are  identical   in  
all the luminosity and stellar mass autocorrelation estimates.

It is important to point out that we have ignored the undersampling of
DEEP2 galaxies on  small scales due to the  slit mask target-selection
algorithm.   \citet{Coil-06a}   use  the  mock   galaxy  catalogue  of
\citet{Yan-White-Coil-04} to  correct for the effect of  this on their
measurements of projected two-point autocorrelation functions, finding
that the corrections are most significant on scales less than $r_p=0.3
h^{-1}$Mpc.    This    scale   is    close   to   the    lower   limit
($r_p\sim0.2h^{-1}$Mpc)   to  which   we   plot  our   autocorrelation
functions. We decided not to attempt to correct for slit mask effects,
but one should  keep in mind that our  $w_p(r_p)$ measurements will be
underestimated  for  $r_p<0.3  h^{-1}$Mpc,  by  roughly  20\%  on  the
smallest scales \citep[see][]{Coil-08a}.

\section{Results}

\subsection{Projected autocorrelation functions and bias factors}

\begin{figure*}
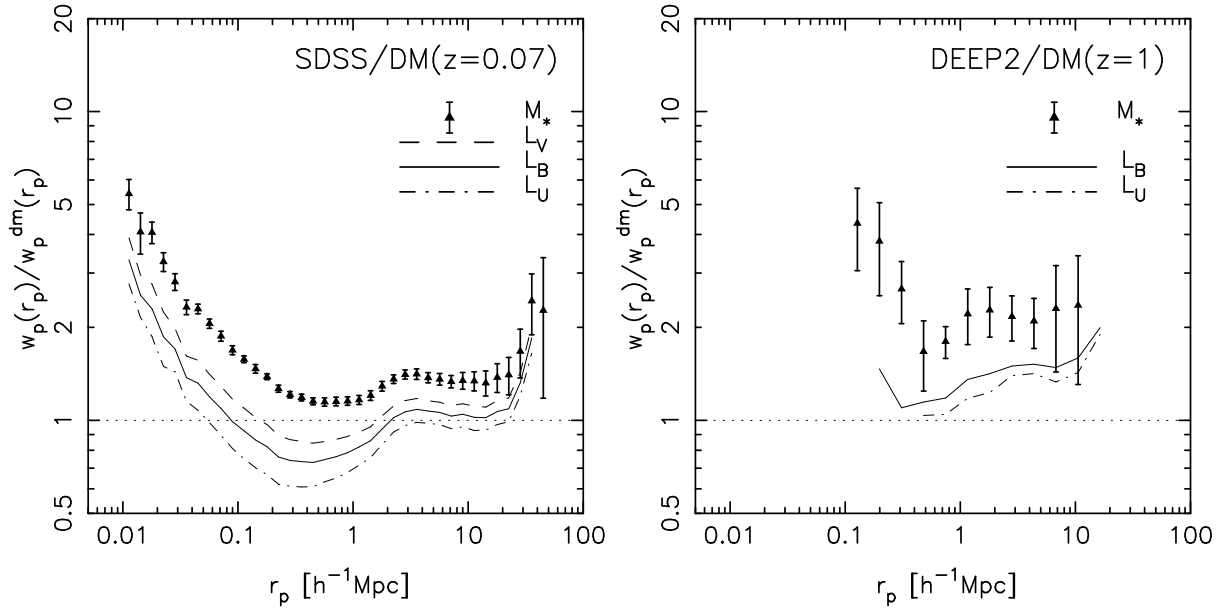

  \centerline{   
    \epsfig{figure=f2a.ps,width=0.45\textwidth}
    \epsfig{figure=f2b.ps,width=0.45\textwidth} }
    \caption{Ratio  of  the  projected  stellar mass  (triangles)  and
      luminosity  (lines) autocorrelation  functions  in the  SDSS/DR7
      (left panel) and in the DEEP2/DR3 (right panel) to the projected
      dark  matter autocorrelation function  at $z=0.07$  (left panel)
      and at $z=1$ (right panel) in the Millennium Simulation.  Errors
      on  the  stellar  mass  autocorrelations  are  estimated  as  in
      Fig.~\ref{fig:wrp}. Errors  on the simulated  functions are much
      smaller.}
   \label{fig:bias}
\end{figure*}

In Figure~\ref{fig:wrp} we show projected stellar mass autocorrelation
functions   $w_p^\ast(r_p)$,   as   well   as   projected   luminosity
autocorrelations $w_p^L(r_p)$ in the $U$, $B$, $V$ bands.  Results are
plotted for  the SDSS  in the  left-hand panel, and for  the DEEP2  in the
right-hand  panel.   In both  panels  $w_p^\ast(r_p)$  is shown  using
triangles  with error  bars, while  the $w_p^L(r_p)$  are  shown using
lines.

The left-hand panel of  Figure~\ref{fig:wrp} reproduces the results of
Paper II where luminosity autocorrelation functions were estimated for
the  five passbands  of SDSS  ($ugriz$). Luminosity  is  less strongly
clustered  than stellar  mass  on  all scales  and  in all  wavebands.
Furthermore, the amplitude and slope of the luminosity autocorrelation
function changes systematically with wavelength, with the reddest band
(the  $V$-band)  showing  the   highest  amplitude  and  the  steepest
slope. As a result, $V$-band  light clusters more similarly to stellar
mass than the bluer bands plotted,  but not as closely as the $z$-band
light analysed in  Paper II.  This is expected  because $z$-band light
is   known  to  be   closely  related   to  stellar   mass  indicators
\citep[e.g.][]{Kauffmann-03}.    Finally,   all  the   autocorrelation
estimates are well described by  power laws. The best power-law fit to
the stellar mass autocorrelations is  shown in the figure, but in this
paper we will  not discuss such fits further, since  they were a major
topic  of the  two previous  papers and  the DEEP2  data do  not allow
autocorrelation shapes to be studied  in as much detail as is possible
at low redshift using SDSS.

As can be seen from  the right-hand panel of Figure~\ref{fig:wrp}, the
DEEP2  galaxies show  very  similar autocorrelation  behaviour to  the
SDSS. Although the measurement errors  are relatively large due to the
smaller sample size  and the considerable depth of  the DEEP2, all the
systematic trends  seen above  for the SDSS  hold also for  the DEEP2.
Furthermore,   the  slopes  of   the  autocorrelation   functions  are
consistent  with  remaining constant  from  $z\sim1$  to the  present,
although  their amplitudes  have increased  by a  factor of  about 1.6
when, as here, they are  measured in comoving coordinates.  To be more
quantitative, we have estimated the relative bias factor as a function
of projected  separation between  the SDSS and  the DEEP2  by directly
comparing their stellar mass autocorrelations.  A linear fit indicates
that the  bias is well described  by $b = 1.6  - 0.09 \log_{10}(r_p)$,
with a reduced $\chi^2$ of 1.03.  The {\it rms} scatter around the fit
is 22.2\%  over the  $r_p$ range probed.   Thus the slopes  of stellar
mass autocorrelation from  the two surveys are really  very similar to
each other. Our result is  broadly consistent with previous studies of
the  evolution of  galaxy two-point  correlation functions  using data
from the same surveys \citep[e.g.][]{Coil-06a,Coil-08a}.

These  results are shown  again in  Figure~\ref{fig:bias} in  terms of
{\em  bias factors}  which we  define as  the ratio  of  the projected
stellar  mass (or  luminosity) autocorrelation  to the  projected dark
matter  density  autocorrelation.  The  latter is  obtained  from  the
$z=0.065$ and  $z=1$ snapshots of  the Millennium Simulation  and thus
assumes the cosmological parameters of the simulation. As noted above,
these redshifts are appropriate to  characterise the mean depth of our
SDSS   and   DEEP2  measurements.   A   maximum  line-of-sight   depth
$|\Delta\pi| = 40 h^{-1}$Mpc  was adopted when computing the projected
autocorrelation  function  for  dark  matter  in order  to  mimic  our
procedures  for the  real  data.  At $r_p\ga  1.5  h^{-1}$Mpc all  our
estimates are  consistent with bias being  scale-independent. On these
scales the  overall bias at $z\sim 1$  is higher by a  factor of $\sim
1.7$ than at $z\sim0$.  This reflects the well-established result that
large-scale structure has been growing more rapidly in the dark matter
distribution since $z=1$ than in the galaxy distribution.

At  smaller separations the  scale-dependence of  the bias  is strong,
with    a     total    range    of     a    factor    of     5    over
$10h^{-1}$kpc$<r_p<1h^{-1}$Mpc  in the SDSS,  and a  factor of  2 over
$100h^{-1}$kpc$<r_p<1h^{-1}$Mpc in the DEEP2.  Another obvious feature
is a {\em  step} at around 2  Mpc which is seen in  all bias functions
and at both redshifts. Considering  the large error bars, we would say
that the step  feature in the DEEP2 bias  function is significant only
at  about  1.5$\sigma$  (at  0.5  $h^{-1}$Mpc) to  2$\sigma$  (at  0.7
$h^{-1}$Mpc) level.   In order  to quantify how  closely the  SDSS and
DEEP2 bias factors approximate each other, we have estimated the ratio
of the two bias functions shown in Figure~\ref{fig:bias} as a function
of projected separation.  This can  also be well described by a linear
function $b = 0.58 +  0.08 \log_{10}(r_p)$, with a reduced $\chi^2$ of
0.4. The {\it rms} scatter around this line is 12.7\%.

Given the featureless $w_p(r_p)$  observed for stellar mass and light,
the strong  scale dependence  of the bias  factors on these  scales is
largely  due to the  much more  pronounced features  seen in  the dark
matter autocorrelations, mainly the  remarkable change in slope at the
transition  between  the one-halo  term,  where  the  pair counts  are
dominated by  galaxy pairs  in the same  halo, and the  two-halo term,
where galaxy  pairs are mostly in  separate haloes, at a  few Mpc (see
figures 2 and 6 of Paper  I for illustrations).  As discussed in Paper
I, this uncomfortable step feature suggests that the amplitude of mass
fluctuations is too  high in the Millennium Simulations,  and that the
shape of our bias functions might be used to estimate the value of the
fluctuation amplitude parameter $\sigma_8$.  We will come back to this
point in the last section.

\subsection{Comparisons with a galaxy formation model}
\label{sec:sam}

\begin{figure*}
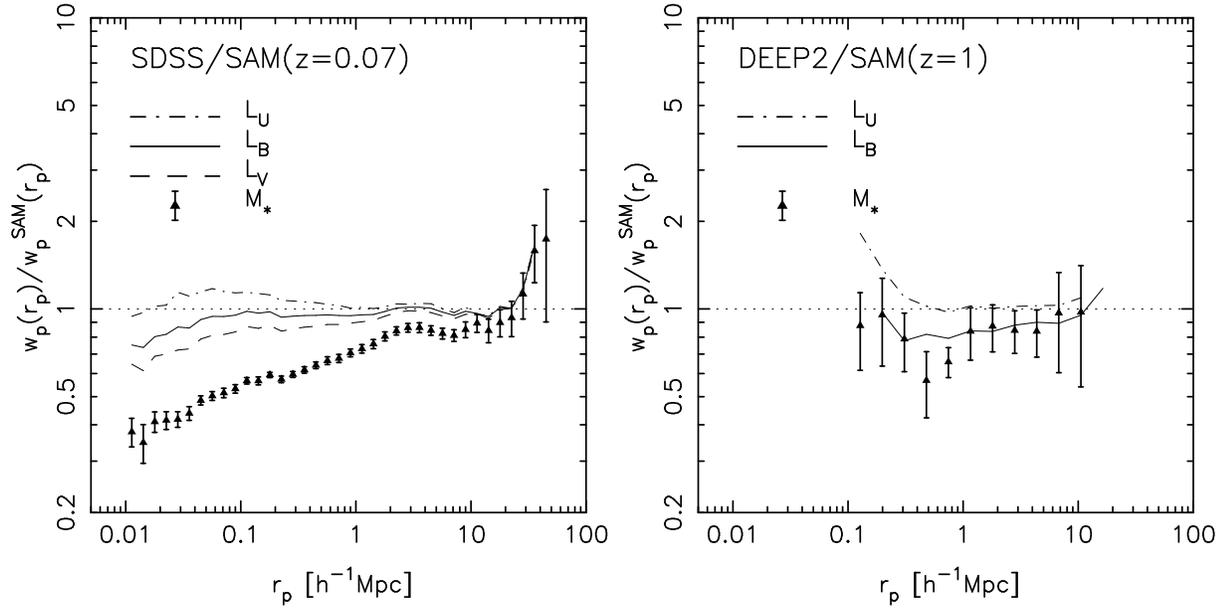

  \centerline{
    \epsfig{figure=f3a.ps,width=0.45\textwidth}
    \epsfig{figure=f3b.ps,width=0.45\textwidth}}
   \caption{Ratio  of  the  projected  stellar  mass  (triangles)  and
     luminosity  (lines)  autocorrelation  functions in  the  SDSS/DR7
     (left panel) and in the DEEP2/DR3 (right panel) to predictions of
     these  same quantities  from the  semi-analytic  galaxy formation
     simulation (SAM)  of \citet{Guo-11} at $z=0.07$  (left panel) and
     at $z=1$ (right panel).}
   \label{fig:wrp_sam}
\end{figure*}

In this section we  compare the observed projected autocorrelations of
stellar light and  mass to predictions for these  same statistics from
the  galaxy  formation model  of  G11,  based  on the  Millennium  and
Millennium-II  simulations.   In  their  paper these  authors  already
compared their  model to SDSS  clustering data, in particular,  to the
projected autocorrelations  of stellar mass and  of galaxies separated
into passive and actively star-forming objects in a series of disjoint
ranges of  total stellar  mass.  They found  their model  to reproduce
observed clustering reasonably well, to a level better than about 20\%
at all separations for $M_\ast\ge6\times10^{10}M_\odot$ and at $r_p>2$
Mpc   for    $M_\ast>6\times10^{9}M_\odot$,   but   to   substantially
overpredict    the   clustering   of    stellar   mass    at   smaller
separations. Further  comparisons between the  SDSS and the  model for
autocorrelation functions  in different intervals of  stellar mass and
optical colour  showed this  discrepancy to be  due mainly  to passive
galaxies      with      stellar      masses     in      the      range
$6\times10^{9}M_\odot<M_\ast<6\times10^{10}M_\odot$,  indicating  that
the model predicts too many red, passive, satellites of this mass
\footnote{The ``excess''  of passive red satellite  is well documented
  in  the literature  and seems  to be  a ``common''  problem  for all
  recent galaxy  formation models.}.  The authors suggested  that this
might indicate  that the fluctuation  amplitude $\sigma_8=0.9$ adopted
in the simulations is too high.

Here we extend this study  to $z=1$ by comparing the model predictions
with    our    DEEP2   results.    The    results    are   shown    in
Figure~\ref{fig:wrp_sam}  (the right-hand  panel), where  we  plot the
ratios  of observed  to  predicted $w_p(r_p)$  both  for stellar  mass
(symbols with error  bars) and for luminosity in the  $U$, $B$ and $V$
bands (lines). For comparison,  the corresponding results for SDSS are
shown  in  the left-hand  panel.  At  both  redshifts the  discrepancy
between   model   and  data   is   clearest   in   the  stellar   mass
autocorrelations.  As  already  found  by \citet{Guo-11},  the  model
$w_p^\ast(r_p)$ exceeds  that observed in  the SDSS on all  scales, by
only  $\sim$15\%  for $r_p\ga2h^{-1}$Mpc  but  by increasingly  larger
factors  on   smaller  scales,  reaching   a  factor  of   $\sim3$  at
10$h^{-1}$kpc.  At   redshift  $z\sim  1$ our  DEEP2  results show  a
similar offset  and are  consistent with the  same trend  within their
error bars.

Remarkably, although the  model predicts stellar mass autocorrelations
that  disagree with  the  SDSS data,  its luminosity  autocorrelations
match  observation   much  better,  particularly   in  the  rest-frame
$B$-band. Apparently  the overly strong clustering of  stellar mass is
almost exactly  compensated by the overabundance  of passive satellite
galaxies so that the light distribution is well reproduced.  At $z\sim
1$ the  predicted luminosity  autocorrelations lie somewhat  above the
DEEP2 measurements  in $B$  but agree well at  $U$.  Given the
error  bars,  the  discrepancy  at  the  longer  wavelengths  is  only
marginally  significant.  The  fact  that a  physically  based  galaxy
formation model can simultaneously  agree with the clustering of light
and disagree with  that of stellar mass, yet  agree with the abundance
of galaxies  as a function  of both light  and stellar mass  (see G11)
demonstrates  the complexity  of the  constraints on  galaxy formation
implied  by  precise observations  of  abundance  and clustering.  The
agreements and  disagreements with  DEEP2 data at  $z\sim 1$  are very
similar to those  with SDSS data at low  redshift, suggesting that the
model  is  treating  the   evolution  and  clustering  of  the  galaxy
population in a realistic way.

\section{Summary and Discussion}

\begin{figure*}
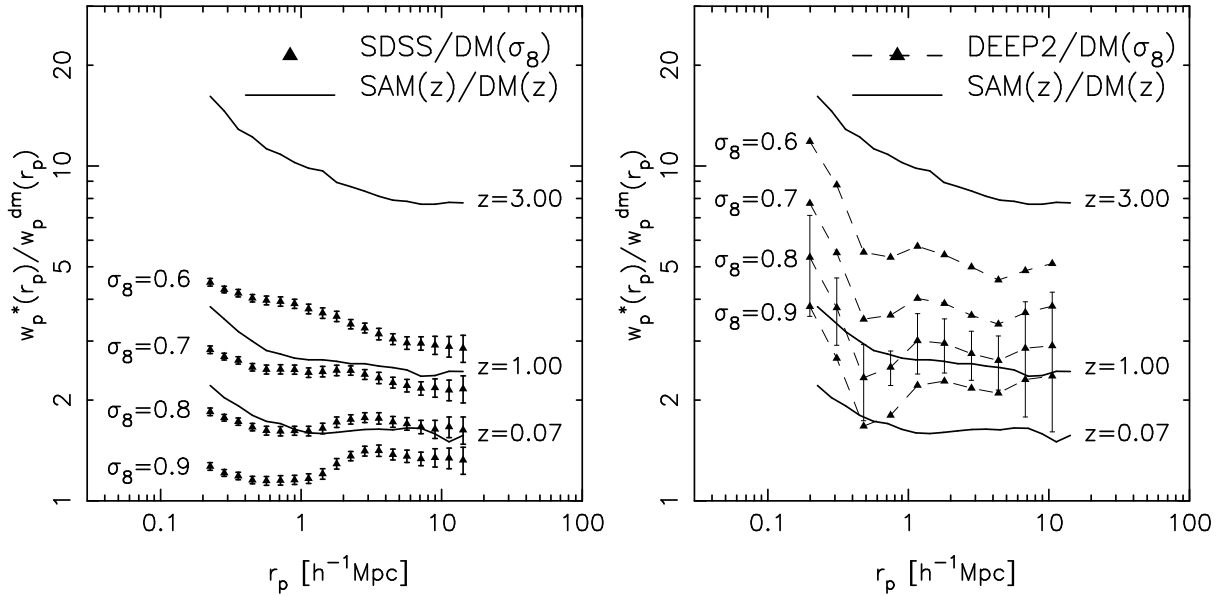

  \centerline{
    \epsfig{figure=f4a.ps,width=0.45\textwidth}
    \epsfig{figure=f4b.ps,width=0.45\textwidth} }
   \caption{Triangles with  error bars indicate bias as  a function of
     scale for stellar mass in the  SDSS (left panel) and in the DEEP2
     (right  panel) assuming the  Millennium Simulation  cosmology but
     with varying values (as labelled) for the present-day fluctuation
     amplitude  $\sigma_8$.   For $\sigma_8$  values  other than  0.9,
     projected dark matter autocorrelations at $z=0.07$ (for the SDSS)
     and $z=1$ (for the DEEP2) are approximated by the functions found
     in  the Millennium Simulation  at earlier  times when  the linear
     fluctuation  amplitude matches that  desired. Bias  functions for
     stellar mass in the  \citet{Guo-11} galaxy formation model (which
     assumes $\sigma_8=0.9$) at  $z=0.07$, 1 and 3 are  shown by solid
     lines, labelled  by redshift.  For  clarity, error bars  are only
     shown in the right-hand panel in the case $\sigma_8=0.8$, and the
     triangles  for   each  $\sigma_8$  are  connected   by  a  dotted
     line.  Only  for  $\sigma_8=0.8$  or  0.7 is  the  bias  function
     inferred from  the data approximately as flat  and featureless at
     the one-halo/two-halo transition as that measured in the model.}
   \label{fig:sigma8}
\end{figure*}

In  this work  we have  estimated projected  autocorrelation functions
$w_p(r_p)$ for the stellar mass of galaxies and for their light in the
$U$, $B$ and $V$ bands, using data both from the third data release of
the DEEP2  Galaxy Redshift  Survey and the  final data release  of the
Sloan Digital Sky Survey (SDSS).   We have compared these to projected
autocorrelations  of   dark  matter  estimated   from  the  Millennium
Simulations at  $z=1$ and $0.07$,  the median redshifts of  our galaxy
samples, in order to investigate the bias of the clustering of stellar
mass and light.

The dependence of clustering on  luminosity and colour for galaxies in
the   DEEP2   was   previously   studied   by   \citet{Coil-06a}   and
\citet{Coil-08a}: the former investigates the luminosity-dependence of
galaxy clustering,  while the latter investigates the  joint color and
luminosity-dependence.   These  studies show  that  the dependence  of
clustering on color is stronger than on luminosity. These findings are
qualitatively very similar to those  seen at low redshift in SDSS data
\citep[e.g.][]{Zehavi-05,Li-06c,Zehavi-11}. Their  conclusion was that
the correlations  between environment and galaxy  properties which are
well known at $z=0$ were, in fact, largely in place at $z=1$.  In this
paper  we  have  investigated   these  issues  further  by  evaluating
alternative  two-point statistics,  the projected  autocorrelations of
stellar mass and  light, in a uniform way for the  SDSS and the DEEP2,
so that the two surveys can  be compared directly to each other and to
the predictions of  a recent MS-based simulation of  the formation and
evolution of  the galaxy population  \citep{Guo-11}.  These statistics
have some advantages  in that they allow the  full observed samples to
be used  to characterize the  distribution of stars at  high precision
over the comoving ranges 10~kpc to  30~Mpc in the SDSS, and 200~kpc to
10~Mpc in DEEP2.

The most impressive  result of this work is  the remarkable similarity
in behaviour  between the  SDSS and the  DEEP2.  The stellar  mass and
stellar light  autocorrelations are, to within  the still considerable
error  bars for  DEEP2, identical  in the  two surveys  except  for an
overall offset in amplitude by about a factor of two.  Papers I and II
have found  that the autocorrelations of stellar  mass (or luminosity)
are dominated  by contributions from  galaxies in a  relatively narrow
range in mass (or luminosity),  around the characteristic value of the
stellar  mass (or  luminosity) function.  In effect,  this  paper thus
compares clustering of the population around $M^\ast$ (or $L^\ast$) at
the two redshifts. The similarity between the two surveys supports the
conclusion of  earlier analyses  that the environmental  dependence of
galaxy properties seen  in the local Universe was  already in place at
$z=1$   \citep[e.g.][]{Coil-06a,   Coil-08a,   Cooper-06,   Meneux-06,
  Meneux-08,   Meneux-09}.    This   similarity,   as  well   as   the
factor-of-two offset between the two redshifts, are well reproduced by
the  galaxy formation  model of  G11 which  matches the  clustering of
stellar  light very well  and overpredicts  the clustering  of stellar
mass to a similar extent at both epochs.

The   galaxy  formation  simulation   fits  the   observed  luminosity
autocorrelations  to within  30\% on  all  scales, both  at $z=0$  and
$z=1$, and it fits the  stellar mass autocorrelations to about 15\% at
$r_p>2$ Mpc.   On smaller scales  the discrepancy in the  stellar mass
autocorrelations reaches  a factor  of 2 to  3. Correlations  on large
scales are produced by galaxies in different haloes (predominantly the
central galaxies  of those haloes)  and the tight  correlation between
central galaxy  properties and  halo mass means  that any  model which
fits the observed abundance of galaxies produces approximately correct
correlations   on    these   scales   \citep[e.g.][]{Vale-Ostriker-04,
  Conroy-Wechsler-Kravtsov-06, Wang-06,  More-09, Moster-10, Guo-10b}.
Smaller scale  correlations result from  galaxy pairs residing  in the
same halo, so the discrepancy on these scales shows that the model has
too many  pairs of  relatively massive galaxies  which share  a common
halo. This may be explained by  too large a value of $\sigma_8$ in the
Millennium  Simulations, which  results in  too many  high-mass haloes
which would then  host these pairs.  The much  better agreement of the
stellar light  autocorrelations on these  same scales, shows  that the
``extra''  satellite galaxies  must have  relatively little  light for
their mass, thus high stellar mass-to-light ratios.

An   interesting   feature  in   the   observed   bias  functions   of
Fig.~\ref{fig:bias}  is  the  obvious  step at  the  one-halo/two-halo
transition at about 2 Mpc which reflects the marked change in slope of
the dark matter correlations at this point (see figs. 2 and 6 of Paper
I). We  noted in Paper I  that since physically  consistent models for
the evolution of the dark  matter and galaxy distributions show smooth
bias behaviour over this  separation range, this feature suggests that
the  amplitude of  mass fluctuations  is  too high  in the  Millennium
Simulation, producing a one-halo/two-halo transition which is stronger
in the simulated mass distribution than in the true mass distribution.
If this interpretation is correct,  the { shape} of our bias functions
can  be  used to  estimate  the  value  of the  fluctuation  amplitude
parameter   $\sigma_8$.     We   illustrate   this    possibility   in
Figure~\ref{fig:sigma8}  where we  plot the  bias determined  from the
clustering   of  stellar   mass   over  the   separation  range   $200
h^{-1}kpc<r_p<15 h^{-1}Mpc$  both for the SDSS ($z=0.07$)  and for the
DEEP2 ($z=1$).  Here we use  dark matter correlations measured  in the
Millennium Simulation not only at the median redshifts of the observed
samples,  but   also  at  higher  redshifts  in   order  to  represent
(approximately) the correlations expected at the observed redshifts in
cosmologies  similar  to  the  Millennium  cosmology  but  with  lower
$\sigma_8$.  We  choose  earlier  redshifts  with  linear  fluctuation
amplitudes corresponding  to $\sigma_8=0.8$,  0.7 and 0.6.  The filled
triangles show estimates of the  stellar mass bias for the two surveys
made in this way for four  different values of $\sigma_8$. It is clear
that  as  the fluctuation  amplitude  is  decreased  the step  feature
becomes less  prominent in  both surveys and  the bias becomes  a more
strongly decreasing function of scale.

Theoretical  predictions for  these  stellar mass  bias functions  for
$z=0.07$, 1 and  3 are shown for the galaxy formation  model of G11 as
solid  lines in  Figure~\ref{fig:sigma8}.  As for  the earlier  galaxy
formation  models plotted in  Paper I,  none of  these curves  shows a
feature at the one-halo/two-halo transition, but their slope increases
significantly   towards   higher    redshift.   Comparing   with   the
observational  results, it  is clear  that the  the best  agreement in
amplitude occurs  for $\sigma_8=0.8$ whereas the  overall shape agrees
best  for  somewhat lower  amplitudes.  It  is  remarkable that  these
results  hold both  for the  SDSS  at $z=0.07$  and for  the DEEP2  at
$z=1$. Hence  the models appear  to describe the overall  evolution of
the  stellar   mass  autocorrelations  surprisingly   well,  once  the
cosmology is  corrected to a  lower $\sigma_8$ value, similar  to that
indicated  by  most  other  recent  analyses of  CMB  and  other  data
\citep[e.g.][]{Komatsu-09,Komatsu-10} (However see \cite{Wang-08a}
who argued that lowering $\sigma_8$ would not always solve the problem.).

In  this paper  we have  only considered  the diagonal  errors  on our
autocorrelation   function  measurements,  since   we  wish   only  to
characterise   the  current   situation,  presenting   a  side-by-side
comparison  between  the SDSS  and  DEEP2  in  terms of  the  observed
autocorrelations, the  bias relative to dark matter,  and a comparison
with the {\em  current} semi-analytic model. To go  beyond this and to
make rigorous model fits to  the full ensemble of $w_p(r_p)$ estimates
would be a major undertaking that is not possible here.

\section*{Acknowledgments}

We are  grateful to  the referee for  his/her helpful comments  on our
paper.   This  work is  sponsored  by  NSFC  (no. 11173045),  Shanghai
Pujiang  Program  (no. 11PJ1411600)  and  the CAS/SAFEA  International
Partnership  Program for Creative  Research Teams  (KJCX2-YW-T23).  CL
would like  to acknowledge the support  of the 100  Talents Program of
Chinese Academy of Sciences (CAS) and the exchange program between Max
Planck Society and CAS. YPJ is supported by NSFC (no. 10821302, 10878001,
11033006).  YPJ and GDL  acknowledge financial  support from the 
European  Research  Council  under  the European  Community's  
Seventh Framework Programme  (FP7/2007- 2013)/ERC grant  agreement n. 
202781. WZ acknowledges financial support from NSFC (no. 10903011).

Funding for the DEEP2 Galaxy Redshift Survey has been provided in part
by NSF grant AST00-71048 and  NASA LTSA grant NNG04GC89G.  Funding for
the  SDSS  and SDSS-II  has  been provided  by  the  Alfred P.   Sloan
Foundation,  the  Participating  Institutions,  the  National  Science
Foundation, the  U.S.  Department of Energy,  the National Aeronautics
and Space Administration, the  Japanese Monbukagakusho, the Max Planck
Society,  and the Higher  Education Funding  Council for  England. The
SDSS Web  Site is  http://www.sdss.org/.  The SDSS  is managed  by the
Astrophysical    Research    Consortium    for    the    Participating
Institutions. The  Participating Institutions are  the American Museum
of  Natural History,  Astrophysical Institute  Potsdam,  University of
Basel,  University  of  Cambridge,  Case Western  Reserve  University,
University of Chicago, Drexel  University, Fermilab, the Institute for
Advanced   Study,  the  Japan   Participation  Group,   Johns  Hopkins
University, the  Joint Institute  for Nuclear Astrophysics,  the Kavli
Institute  for   Particle  Astrophysics  and   Cosmology,  the  Korean
Scientist Group, the Chinese  Academy of Sciences (LAMOST), Los Alamos
National  Laboratory, the  Max-Planck-Institute for  Astronomy (MPIA),
the  Max-Planck-Institute  for Astrophysics  (MPA),  New Mexico  State
University,   Ohio  State   University,   University  of   Pittsburgh,
University  of  Portsmouth, Princeton  University,  the United  States
Naval Observatory, and the University of Washington.

\bibliography{ref}

\bsp
\label{lastpage}

\end{document}